# Engaging Software Engineering Students in Grading

## The effects of peer assessment on self-evaluation, motivation, and study time

Wouter Groeneveld, Joost Vennekens, Kris Aerts
OVI and LESEC, Department of Computer Science, KU Leuven
Leuven, Belgium
{firstname}.{lastname}@kuleuven.be

## ABSTRACT

Peer assessment is a popular technique for a more fine-grained evaluation of individual students in group projects. Its effect on the evaluation is well studied. However, its effects on the learning abilities of students are often overlooked. In this paper, we explore self-evaluation, motivation, and study time of students in relation to peer assessment, as part of an ongoing project at our local Faculty of Engineering Technology. The aggregated measurements of two years so far show that: (1) students get much better at evaluating their own project on some, but not all, of the evaluation criteria after a peer assessment session, (2) students report in a follow-up survey that they are more motivated to work on their project, and (3) the relation between motivation and time spent on the project increases. These results suggest that peer grading could have positive long-term effects on the reflective, and therefore lifelong learning, skills of students. A better understanding of the evaluation criteria results in more accurate self and peer grades, emphasizing the importance of properly defining and communicating these criteria throughout the semester.

## 1. INTRODUCTION

Engineering students are faced with the challenging task of gaining technical knowledge in their specific field to be able to graduate and hopefully cope with future professional environments in the industry. However, technical expertise does not suffice: several non-technical skills such as self-reflection, lifelong learning, creativity, and empathy are also expected of freshly graduated engineers, whether it is in the field of software engineering [1] or in any other engineering field [2].

In order to enhance the learning of these non-technical skills, without making any compromises to the gaining of technical knowledge, we propose the introduction of peer assessment. Peer assessment has often been used to enhance exactly these skills [3,4]. The development of self-evaluative skills and the awareness that lifelong learning is a necessity for an engineer are our main goals. We are therefore interested in the following research questions, for which there is currently only sporadic evidence:

- **Q1**: *What are the effects of peer assessment on self-evaluation?*
- **Q2**: *What are the effects of peer assessment on motivation and study time?*

The remainder of this paper is divided into the following sections. Section 2 describes background information and related work on peer assessment, and why it is of growing importance. Section 3 clarifies the process we have followed to collect data of two years. Next, in Section 4 we present and discuss our findings. Possible limitations are identified in Section 5, while Section 6 concludes this work.

## 2. BACKGROUND AND RELATED WORK

As stated by Sitthiworachart and Joy, "*The use of peer assessment is claimed to enhance students' evaluative capacities, which improve the quality of their subsequent work*" [5]. Several papers, even from more than 20 years ago, argued that using self and peer assessment increases the students learning capabilities [4]. Peer assessment is indeed "*as much about learning as assessment*" [5]. However, many papers lack a good body of empirical data to back up these claims, except for qualitative surveys and self-reports.

It is important to note that in the context of this work, we are in not focusing on the validity of peer assessment as a tool to completely replace the grading of teachers. According to Sajjadi et al., intricate models that would correct any bias or variance in students' grading seems to be rather ineffective [6]. Researchers do not completely agree on the validity of peer grading: some find it to be adequate [7,8], while others find it very variable [9,10]. As our intention is to introduce peer assessment primarily as a learning tool for students, teacher assessment was not completely removed. Section 3 explains this grading system in greater detail.

## 3. METHODOLOGY

In a 'Software Design in C++' course for third year bachelor students Electronics/ICT at our local Faculty of Engineering Technology, we asked students to grade themselves, fill in a follow-up survey on motivation, and to keep track of their study time on the project of the course. In the second academic year, peer assessment was introduced.

The first academic year, 2018-19 (N = 25 students), served as a baseline, while the introduction of peer assessment in the second academic year, 2019-20 (N = 27 students), served as the study group. As part of the evaluation process for this course, students are required to complete an integrated assignment presented as a project, which accounts for `50%` of the total score (divided into `27%` peer grades, `23%` teacher grades). Project assignments are very open-ended: students can choose to implement anything they like, as long as the following requirements are met:

- The target programming language is `C++11`, and the target platform is the Game Boy Advance. Naturally, every student opts for the creation of some kind of 2D game.
- A GBA sprite engine was custom built for this course to reduce the technical challenges of working with a low-level embedded hardware system such as the GBA [11]. Students were required to start from this engine.

An evaluation rubric was adopted from [10] to make both the self and peer evaluation process easier for students. This rubric was also employed by the teaching staff in both academic years. Since the Software Design course is a programming course, different non-technical and technical criteria were needed to evaluate different aspects of student projects. Table 1 summarizes this rubric and provides examples of low and high scores.

Cardinal rating was employed by providing a score for each criterion between 0 to 5, after which weights are assigned and a global score on 20 is calculated.

Table 1: The qualitative evaluation rubric used to grade the projects.

| Weight | Criteria | Examples |
|---|---|---|
| 0.4 | **Code Design** | *How well-designed is the project code?* High score: clearly recognizable objects, represented in domain model, separation of concerns. Low score: All code in single object, unclear what does what, barely or no model. |
| 0.5 | **Clean Code** | *How readable is the project code?* High score: use of understandable variables, methods, classes. Low score: Too much re-reading is needed to see what is happening. |
| 0.4 | **Complexity** | *How difficult was the project made?* High score: chosen for a challenge instead of a simple implementation. Low score: path with smallest resistance taken, the bar set too low. |
| 0.3 | **Creativity** | *How original and creative is the project idea?* High score: implemented an original idea instead of a clone of a default 2D platformer. Low score: opted for a less inspiring design, everything is based on existing work |
| 0.2 | **GBA UI** | *How elaborate is the presentation of the game?* High score: All UI/Sound techniques applied well: sprites, scrolling BG, … Low score: little to no animation/backgrounds, monotonous design. |

This rubric is a condensed subset of the learning outcomes of the course. The rubric was made public in the very first lesson so that students could take this into account during the development of their project. Students were encouraged to work together in small groups to further reduce the complexity and stress of cross-compiling for the GBA. Therefore, the rubric was used to individually assess *projects* (and thus, a *group*).

Peer assessment was obligatory and took place a day after the submission deadline. Every group had exactly five minutes to demonstrate their project to their fellow students and the teaching staff. We recommended students to spend at least two minutes explaining the overall structure of the code, since two criteria are technical and evaluate the code quality. After the peer assessments, students were asked to individually evaluate their own project using the very same rubric.

To measure accuracy and motivation, a short follow-up survey was filled in after the self-assessment assignment, in which students had to answer the following questions:

1. How easy was it for you to evaluate yourself? (Likert scale, 1-5)
2. How accurate do you think your evaluation is? (Likert scale, 1-5)
3. How motivated were you while working on the project? (Likert scale, 1-5)
4. What motivated you? (Open-ended question)
5. What demotivated you? (Open-ended question)

To measure study time, students were required to keep track of the amount of time spent on working on the project. A template `csv` file was provided for them to fill in, noting the date and the number of hours spent working on the project that day. A simple `csv` file was used to reduce the administrative load as much as possible, while still maintaining a high accuracy rate. A data correction step was necessary as many students made mistakes in noting the year (1018 instead of 2018, 2018 instead of 2019 in January, …).

Data was not collected anonymously since we wanted to compare self-grading with the actual grading marks. However, since the authors are also the teachers of the course, the data was made anonymous during the analysis of this research to prevent any bias that result from knowing student names.

## 4. RESULTS AND DISCUSSION

The collected data from two years was analyzed to help us estimate the effects of peer assessment on multiple variables. We will discuss the results based on the two research questions from Section 1.

### Q1 - The effects of peer assessment on self-evaluation

To assess the accuracy of self-evaluation, as done in [10], correlations between self and teacher assessment were investigated and shown in Table 2 and in Figure 1. The Pearson correlation coefficient was calculated using the individual grades as elaborated in Section 3. Creativity was added to the evaluation rubric in the second year.

*Table 2: Comparison of self (Stu.) and teacher assessment (Tea.): mean, standard deviation, and correlation values for each criterion in the evaluation rubric.*

|  |  | Design | | Clean Code | | Complexity | | UI | | Creativity | |
|---|---|---|---|---|---|---|---|---|---|---|---|
|  |  | Stu. | Tea. | Stu. | Tea. | Stu. | Tea. | Stu. | Tea. | Stu. | Tea. |
| Year 1 | Mean | 3.17 | 3.71 | 3.54 | 3.08 | 3.13 | 3.04 | 3.21 | 3.50 | / | / |
| | s.d. | 1.010 | 1.197 | 0.658 | 0.881 | 0.850 | 1.233 | 1.062 | 1.142 | / | / |
| | Corr. | **0.331** | | **-0.231** | | **0.202** | | **0.376** | | / | |
| Year 2 | Mean | 4.10 | 3.33 | 3.98 | 3.00 | 4.13 | 2.96 | 2.67 | 3.19 | 3,67 | 3.31 |
| | s.d. | 0.775 | 0.582 | 0.608 | 1.158 | 0.558 | 1.523 | 0.836 | 1.184 | 0.679 | 0.664 |
| | Corr. | **0.105** | | **0.341** | | **0.666** | | **0.713** | | **0.528** | |

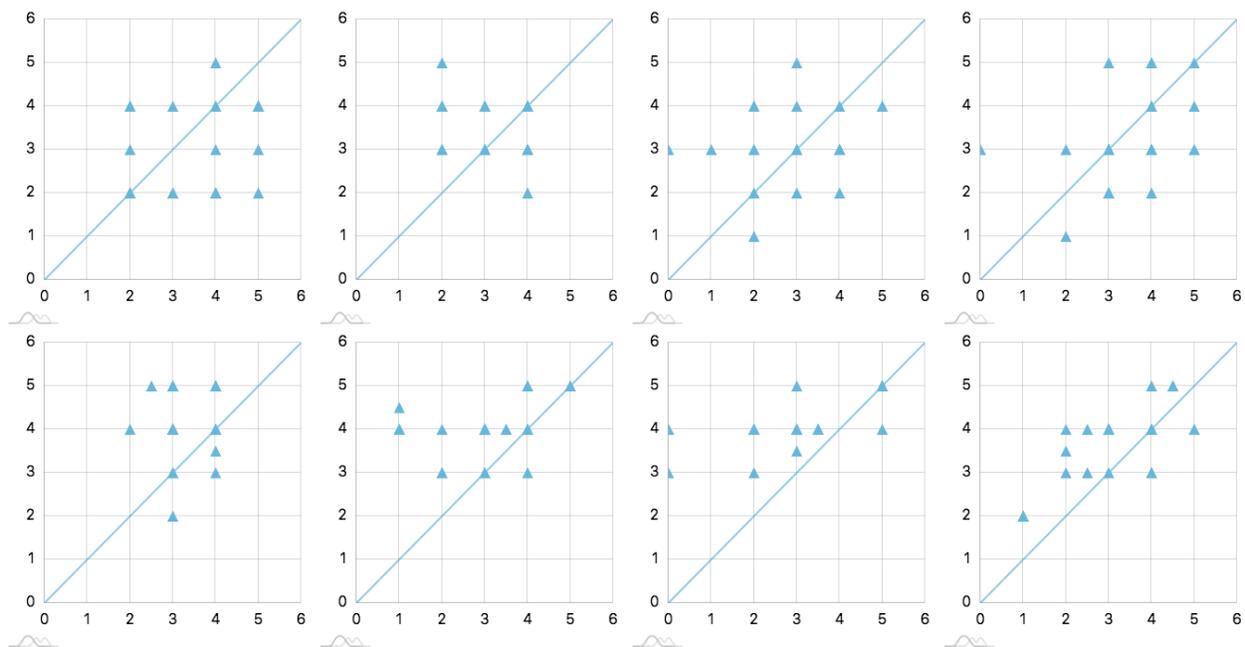

*Figure 1: Scatter plots of self-evaluation (Y-axis) related to teacher-evaluation (X-axis) for each criterion. The closer to the centerline, the higher the correlation and the better the student's estimate. Left to right: Design, Clean Code, Complexity, UI. (Above: year 1. Below: year 2.)*

When comparing the means between different years, we can clearly see that students who participated in peer grading generally graded themselves higher. This bias towards higher student gradings, as reported by other researchers [10], is also visible in Figure 1, where triangles evolve towards the left side of the centerline. Furthermore, as the correlation increases, the deviation decreases, revealing a higher level of agreement among the scores given by different students to the same project.

Students in the second year report that they found it easier to self-evaluate. Means from the Likert scales (1-5) are displayed in Table 3. This could signify that students get more comfortable evaluating projects during the peer grading process.

Table 3: Self-reported means for ease of grading and accuracy. 1 (hard/inaccurate) - 5 (easy/accurate).

| Year | How easy to grade? | How accurate? (reported) |
|---|---|---|
| 2018-19 | 3.25 (*s.d.* 1.073) | 3.25 (*s.d.* 0.608) |
| 2019-20 | 3.42 (*s.d.* 0.758) | 3.62 (*s.d.* 0.571) |

**Evaluating *Design* and *Clean Code***

Students reported in the follow-up survey that it was very hard for them to objectively evaluate the criteria '*Design*' and '*Clean Code*'. One needs a firm grasp of programming techniques and a lot of experience in reading and evaluating code to be able to make a correct judgment. A few lab sessions are effectively dedicated to introducing students to these concepts. However, as they are still inexperienced, it is indeed very difficult to make a good estimate - especially within the a short time-frame of the presentation. No code was assessed in advance.

There is no strong relation between self and teacher grading in *Design* (`0.331` to `0.105`). However, the negative correlation in *Clean Code* evolves into a weak relation (`-0.231` to `0.341`), indicating that students at least have some idea on what is 'clean' and what is not, after they have seen more code examples from their peers. This is an important lesson for us, as in the coming years more examples should be introduced in the labs, where room for discussions can lead to a better understanding of the concept.

Since the evaluation of both technical criteria relies on the assessment of source code text, other assessment methods such as comparative judgement could further increase the reliability of self and peer grades, as discovered by Goossens and De Maeyer [12].

**Evaluating *Complexity* and *UI***

A strong increase in the correlation between self and teacher grading was registered for both *Complexity* (`0.202` to `0.666`) and *UI* (`0.376` to `0.713`), as also visible in Figure 1. The more examples students see and have to grade, the better they can estimate their own project on these criteria. Both *Complexity* and *UI* can be graded by looking at the end product (the GBA game) rather than the source code (C++ files). It is easier to evaluate whether your own game is less or more complex than others if you saw what your peers made. The same holds true for *UI*, where the biggest correlation was noted.

Students frequently applauded others for their ingenuity and resourcefulness. For example, multiple students were asking technical questions on how a group implemented dynamic background switching of their 'Mr. Driller' game. Students got inspired by taking a look at the work of others. However, the group presentation influenced the way the students do the assessment. The correlations could indicate that an introspective process was triggered that benefits the learning process of students.

### Evaluating *Creativity*

Every year, students report that the single biggest motivator for them is the possibility to be creative with the GBA. Therefore, the *Creativity* criterion was added during the second year. No comparisons can be made. However, a weak correlation was registered after the peer grading process (0.528). The problem with creativity is that it can be interpreted in multiple ways, even if examples of good and bad grades have been provided in the rubric.

We are positive that this correlation can be easily increased provided that a clear definition of creativity is given. The projects are only one dimension of the 4P Creativity Model: the Product [13]. The Process dimension is something we already touch upon in the labs by introducing creative techniques to cope with the hardware limitations of the older GBA system. Our aim for the future is to better educate students on what it means to be creative. We are convinced that peer assessment can play a big role in the creativity of students, as also noted by Papaleontiou et al. [14]:

> *[…] As a guide for creativity, in order to promote creativity as product, with a view at the practical site of teaching, teachers should: […] encourage self and peer assessment and evaluation.*

### Q2 - The effects of peer assessment on motivation and study time

### Self-reported motivation

Students report a mean motivation of 3.58 (*s.d.* 0.659) in year 1, and of 3.96 (*s.d.* 0.774) in year 2. It is hard to say that this increase in motivation is due to the introduction of peer grading alone. Therefore, we also asked students to write down their motivators and demotivators, of which a selection of the answers is presented in Table 4.

*Table 4: A sample of the reported motivators and demotivators for the GBA project of the course, throughout both academic years.*

| Stud. | Motivator (+) | Demotivator (-) |
|---|---|---|
| 1 | Creating something out of nothing | Things that did not work on the first few tries such as backgrounds and pointers |
| 2 | Great freedom to choose what is made | reverse engineering the GBA hardware |
| 3 | Passing the course | GBA memory issues |
| 4 | Using own inspiration, open nature of the project | no debugging |
| 5 | Creating a game, gaming to test it | Background difficulties |
| 6 | Being creative and looking for solutions | Bugs |
| 7 | GBA Nostalgia | Exams |
| 8 | It is something else than a default assignment | Limited memory capabilities, complexity |
| 9 | Seeing graphical progress when developing the game | Too much new concepts to learn |

It is interesting to see that none of the students actually mentioned the peer evaluation process itself. However, that does not mean that their motivation was not influenced by the introduction of it. A classic mix of intrinsic ('*fun*') and extrinsic ('*good grades*') motivation was reported, as expected, with a clear bias towards intrinsic motivation. Of course, the fact that one could program on a gaming hardware device is one of the biggest reasons to be highly motivated. The main demotivators are related to the technical challenges of the project assignment.

**Study time measurement breakdown**

Figure 2 visualizes the total invested time per day for all students during the given academic years, divided into three parts. The biggest peak in the first part, mid December, is the last lab where students are allowed to work freely on their project. Most small dips are Sundays. During the second part, the Christmas holiday period, students clearly start worrying about their project and gradually invest more time. In the third part, a race towards the submission deadline begins, ending with almost 100 collective hours on a single day, averaging on 4 hours per student.

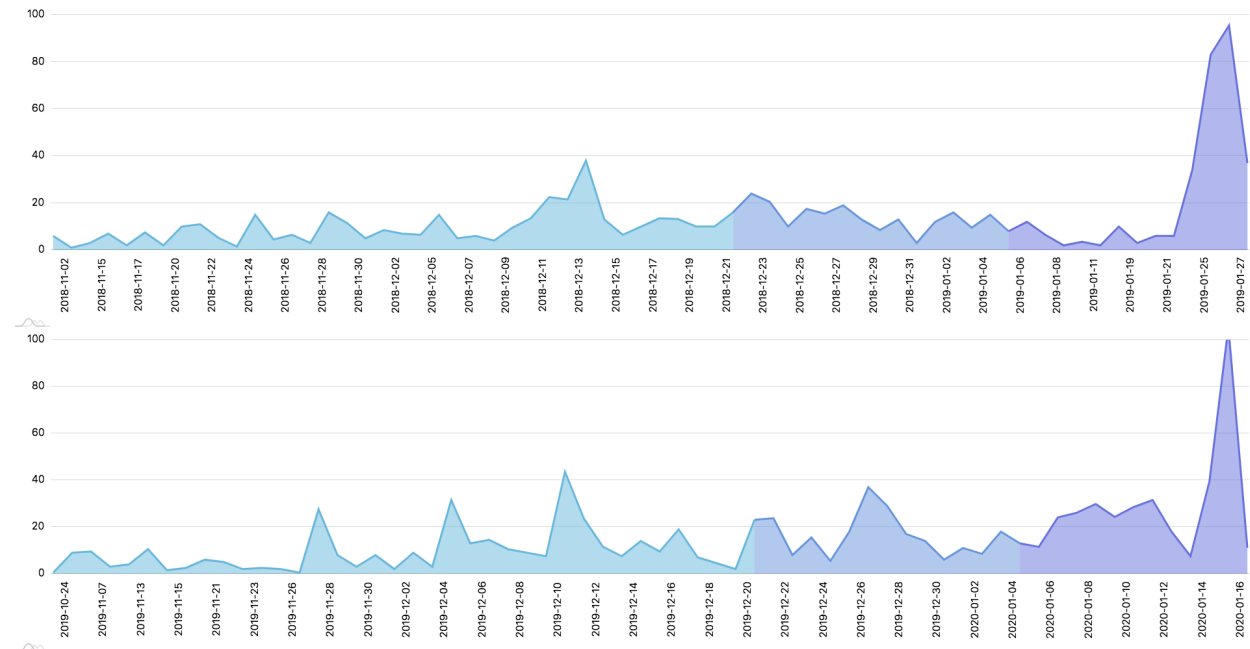

*Figure 2: A normalized time graph of total invested hours, broken down in three color-coded parts: before Christmas break, Christmas break, before deadline. (Above: year 2018-19. Below: year 2019-20.)*

No significant change in workload visible between the two years, except for the overall increase in the last race for the deadline. However, the average amount of hours decreases slightly, being 39.63 (*s.d.* 18.999) in the first year and 36.61 (*s.d.* 18.892) in the second. This leads us to conclude that having to evaluate fellow students does not drastically change the way students take on a project in general. They might overly rely on the good-will of others to give high grades. The usage of the peer evaluation system was stressed multiple times throughout the semester to make sure students remember how the evaluation of the course works, with a small negative effect on study time.

**Link between motivation and study time**

Table 5 and 6 display the correlation matrices for dependent variables time spent, motivation, and total grade. We suspected that motivation and study time investment would be closely linked together. However, without peer grading, no relation could be determined between time spent and motivation. The raw data reveals students with high grades that made little investment into the project, and students with low grades that made big investments. Averages on total grades are 13.73 for year 2018-19, and 13.29 for year 2019-20.

Table 5: Correlation matrix for year 2018-19

|  | Time spent | Motivation | Grade |
|---|---|---|---|
| Time Spent | - |  |  |
| Motivation | 0.005 | - |  |
| Total Grade | 0.394 | -0.367 | - |

Table 6: Correlation matrix for year 2019-20

|  | Time spent | Motivation | Grade |
|---|---|---|---|
| Time Spent | - |  |  |
| Motivation | 0.331 | - |  |
| Total Grade | 0.292 | 0.423 | - |

No strong association could be derived with a correlation higher than 0.5. However, note again the big increase in correlations on motivation linked to time spent and total grade. Coupled with the fact that student's self-reported motivation increases overall, the higher correlation suggests that the peer evaluation caused at least the motivated students to invest more time in the project. In a study carried out by Krapp [15], the distinction was made between intrinsic motivation called interest, and extrinsic motivation. Krapp discovered, as one would expect, that interest is more highly related to study time than extrinsic motivation. Our follow-up survey reveals the presence of both motivator types, which makes it harder get a distinct correlation.

## 5. LIMITATIONS

The biggest threat to the validity of this research is the limited scale of execution. It is not the student group size that is a concern, but the longitude of the research. As we measured only two years to date, more years could mitigate possible shortcomings in our study. However, the strongly deviating correlations presented in Section 4 are enough to substantiate the results. Existing research further supports this. Our intention is to continue with the data collection and the peer grading process in the coming years.

The peer evaluation process itself required students to fill in a form on paper. It is not impossible for students to copy grades of their neighbors while assigning a grade. However, this possible bias and/or variance was mitigated by still taking teacher grades into account. We did not aim for perfect peer grades, but rather for the positive effects of the grading process itself. Extreme positive biases, for example bulk grading or complete random grading, were not encountered. As students could already be motivated for this course, ceiling effects could have occurred on the motivational gains.

## 6. CONCLUSION

Peer assessment drastically improves the quality of self-assessment. Gracias and Garcia ask '*Can we trust peer grading?*' and conclude with mixed results [10]. We would like to change the question to '*Can we trust the positive effects of peer grading?*'. The answer is undoubtedly yes. Students are more motivated and can better evaluate their own work, even if there is no bigger quantifiable time investment measured. In contrast to a report by Sherrard et al. [16], our students did not express any reservations regarding the fairness of the peer grading system. Perhaps this is because teacher grades are still taken into account.

Future work involves examining how peer assessment can help with the development of engineering skills. The combination of peer assessment and creativity deserves special attention. After seeing the time investment interval from Figure 2, we also ask ourselves how to distribute this more proportionally. For instance, earlier and more frequent feedback during the development of the projects could help.